\documentclass[conference,a4paper]{IEEEtran}
\usepackage[]{algorithm2e}

\usepackage{cite}

\ifCLASSINFOpdf
  \usepackage[pdftex]{graphicx}
  \graphicspath{{../pdf/}{../jpeg/}{./figures/}}
  \DeclareGraphicsExtensions{.pdf,.jpeg,.png}
\else
\fi
\usepackage{amsmath}
\usepackage{algorithmic}
\ifCLASSOPTIONcompsoc
 \usepackage[caption=false,font=normalsize,labelfont=sf,textfont=sf]{subfig}
\else
 \usepackage[caption=false,font=footnotesize]{subfig}
\fi
\begin{document}
\bstctlcite{IEEEexample:BSTcontrol}
\title{Exploring a Double Full-Stack Communications-Enabled Architecture for Multi-Core Quantum Computers}

\author{\IEEEauthorblockN{Santiago Rodrigo, Sergi Abadal, Eduard Alarc\'on}\\
\IEEEauthorblockA{NaNoNetworking Center in Catalonia\\
Universitat Polit\`ecnica de Catalunya\\
\{srodrigo, abadal\}@ac.upc.edu, eduard.alarcon@upc.edu}\\

\and

\IEEEauthorblockN{Carmen G. Almudever}\\
\IEEEauthorblockA{QuTech\\
Delft University of Technology\\
C.GarciaAlmudever-1@tudelft.nl}}


%


\IEEEoverridecommandlockouts
\IEEEpubid{\makebox[\columnwidth]{978-1-7281-9132-4/20/\$31.00~\copyright2020 IEEE \hfill} \hspace{\columnsep}\makebox[\columnwidth]{ }}

\maketitle


\thispagestyle{plain}
\pagestyle{plain}

\begin{abstract}
Being a very promising technology, with impressive advances in the recent years, it is still unclear how quantum computing will scale to satisfy the requirements of its most powerful applications. Although continued progress in the fabrication and control of qubits is required, quantum computing scalability will depend as well on a comprehensive architectural design considering a multi-core approach as an alternative to the traditional monolithic version, hence including a communications perspective. However, this goes beyond introducing mere interconnects. Rather, it implies consolidating the full communications stack in the quantum computer architecture. In this paper, we propose a double full-stack architecture encompassing quantum computation and quantum communications, which we use to address the monolithic \textit{versus} multi-core question with a structured design methodology. For that, we revisit the different quantum computing layers to capture and model their essence by highlighting the open design variables and performance metrics. Using behavioral models and actual measurements from existing quantum computers, the results of simulations suggest that multi-core architectures may effectively unleash the full quantum computer potential. 
\end{abstract}


%
\IEEEpeerreviewmaketitle

\newboolean{comms_saw}

\setboolean{comms_saw}{true}

\section{Introduction}


\begin{figure}
    \centering
    \includegraphics[width=0.8\linewidth]{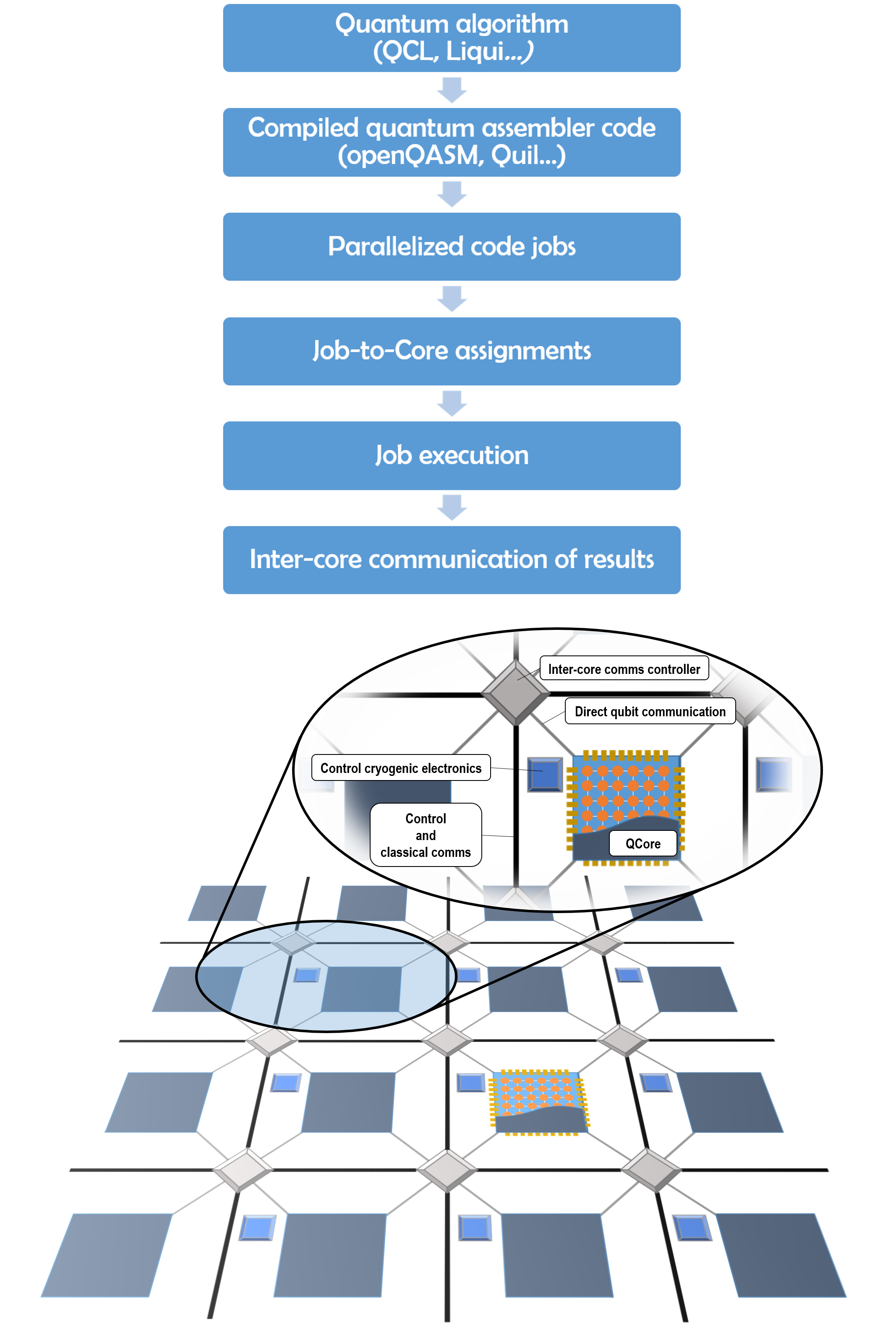}
    \caption{Multi-core quantum computer and code flow}
    \label{fig::multi-core_computer}
\end{figure}



Research on quantum computation forged since the 80s has created considerable expectations on its unprecedented processing power and unconditional security, which could change forever key areas such as cryptography, big data analysis, AI, simulations for physics, material science and biochemistry (drug synthesis), etc. \cite{resch2019quantum,martonosi2019next,wehner2018quantum,ladd2010quantum}. By leveraging quantum mechanical properties such as superposition and entanglement, quantum computer implementations of time-consuming algorithms can be exponentially faster than their classical counterparts. For instance, Shor's algorithm\cite{shor1999polynomial} allows factoring large numbers in polynomial time, whereas no efficient non-quantum algorithm is known, that is, while a classical computer would take a million years to factor a standard 1024-bit RSA key, a quantum computer could do the same task in hundreds of seconds \cite{van2013blueprint}.

However, preserving qubit entanglement and quantum state superposition --the key enablers of the quantum processing power-- implies maintaining the quantum information in qubits (the \textit{alter ego} of classical bits in the quantum world) intact, i.e. keeping information coherence. This, being trivial in classical computing, is in fact one of the most challenging issues for building quantum computers, that is, any interaction with other particles or forces causes a qubit to rapidly decohere and lose the information it contains. Therefore, quantum processors must be kept at very low temperatures (close to the absolute zero) and isolated from the outside world, something which makes the external control and computation, for operations on the qubits and measurements of their values, a very challenging task. 

Although during these last years we have seen remarkable sustained advances on quality and number of qubits in working prototypes, the existing realizations of quantum computers are too small-scale and error-prone yet to be able to experimentally demonstrate the theoretical results and proven algorithms that show these impressive speed-ups \cite{national2019quantum}. As an example, the largest functional quantum processors hitherto presented (IBM and Google have most recently shown both 53-qubit processors at work \cite{IBM2019qcomp53qubits,arute2019quantum}) have several orders of magnitude less qubits than the billion of them needed to factor a 1024-bit integer\cite{preskill2012quantum,preskill2019supremacy}. Actually, there is a fair amount of legitimate skepticism\cite{vardi2019quantum,dyakonov2020will} on whether quantum computing will satisfy the performance, resource and cost requirements for its commercial adoption, despite the dramatic progress of the last decade. In fact, current approach for designing and building quantum computers, based on densely integrating qubits on a single chip, is conjectured not to scale past some hundreds of qubits, due to impracticality of control circuits integration, per-qubit wiring, prohibitive quantum decoherence and severe qubit operation errors\cite{baker2020time,national2019quantum}.

Naturally, work undertaken in the fields of physics, materials, electronic and quantum engineering allows us to expect improvements on the quality and number of qubits integrated in operational quantum processors in the near future. Much of the research being carried out now on scalability of quantum computing aims at extracting the most performance out of the current Noisy Intermmediate-Scale Quantum (NISQ) computers\footnote{NISQ is a term coined by John Preskill that encompasses the small-sized and constrained (yet fascinating) computers built nowadays \cite{preskill2018quantum}} by optimizing compilers \cite{khammassi2020openql,paler2018nisq,chong2017programming,wallman2016noise,javadiabhari2015scaffcc}, qubit mapping and routing \cite{lao2019timing,li2019tackling,tannu2019not,cowtan2019qubit,zulehner2018efficient,lao2018mapping,siraichi2018qubit}, and very importantly, Quantum Error Correction (QEC), i.e. techniques developed for qubit error detection and correction \cite{bennett1996mixed,gottesman2010introduction,cory1998experimental}. Closer to the physics are the works on improving control wiring, and signalling and circuits that may work at near-zero temperatures \cite{isailovic2004datapath,li2018crossbar,patra2017cryo,hornibrook2015cryogenic,homulle2017reconfigurable}. Moreover, several technologies for implementing qubits have been proposed\cite{cirac1995quantum,loss1998quantum,kane1998silicon,nakamura1999coherent,prawer2008diamond}.

However, the \textit{quantum leap} --doubly quantum, we could say-- there exists from today's prototypes to fully-functional and useful quantum computers has such a breadth and depth as to require additional support from other disciplines related to processor design. 
Such approach to the problem demands for a system-wide optimization, the foundations of which may be laid on audacious proposals for the design and architecture of the quantum computer as a whole. 


We postulate that, even though the challenges are hard and diverse, a comprehensive approach of the computer design based on multi-core architectures, as opposed to current densely-packed monolithic approach, is crucial to unlock the scalability issues. This multi-core quantum computer, presented in Fig. \ref{fig::multi-core_computer}, will cluster together dozens of NISQ cores (with tens to hundreds of qubits), connected through a quantum communications network (for core-to-core qubit transport, such as quantum teleportation or photonic switches) and a control classical network (for core coordination and job distribution), mapping the quantum algorithm among them to boost performance. In this way, we alleviate the requirements for control circuits and improve qubit isolation, while leveraging all the advantages of quantum parallelism.

Various proposals \cite{monroe2014large,caleffi2018quantum,brown2016co,vandersypen2017interfacing,jiang2007distributed,sargaran2019saqip,isailovic2006interconnection} in the existing literature agree on using systems based on the interconnection of multiple NISQ quantum processors, in a many-core computing fashion. Existing articles use different qubit technologies (ion trap, quantum dots or impurities in solids) and \textit{module} interconnects (ion shuttling, photonic switches, quantum teleportation), but to the best of our knowledge none of them has deeply analyzed whether this multi-core approach is effectively enabling an architecturally scalable quantum computer, and which are the resource overheads and computational costs of such architectures. Moreover, it's not clear which qubit type or communication technology will behave best.


We aim at substantiating that quantum computing scalability (and ultimately, quantum computing culmination) may not be possible without multi-core architectures enabled by communications, in a Quantum Network on Chip (QNoC) fashion. We do so by performing a first analysis of this approach focusing on the intermodule (or interchip) communication, looking for well-sustained conclusions on whether it allows quantum computers to scale with the number of qubits or the communication process becomes in fact a new bottleneck for the processing power.

In this article, we set the framework for this analysis. We propose a double full-stack layered architecture combining communications with traditional quantum computer designs, and present a Design Space Exploration (DSE) formulation to this problem. DSE will be used to compare the multi-core and monolithic single-core approaches and find a sweet spot (or region of the design space) where the design performs better by examining the parameter space (i.e. qubit technology, qubit and core interconnects, number of cores...) and evaluating it with performance metrics. In this way, it will be possible to adequately detect how the different elements in the system relate to each other and the way each one affects to the overall performance of the system.

The paper is structured as follows. In Section \ref{section::statement} the double-stack communications-enabled layered architecture for quantum computing is presented, whereas Section \ref{section::vars} contains the description of the main variables characterizing the system. Section \ref{section::DSE} is devoted to explaining DSE and its application to our problem, with a brief state-of-the-question on performance metrics for quantum computing. The results of the first scalability exploration and Quantum Technology Gap Analysis are shown in Section \ref{section::results} before drawing some conclusions and setting the future work. 

\section{Connecting the (quantum) dots}
\label{section::statement}


In this section, a generic multi-core quantum computer architecture is presented in order to facilitate the context for the analysis previous to the DSE, which will identify the parameters and performance metrics that best fit our problem.

Some layered architectures for quantum computing have already been proposed \cite{fu2017experimental,van2016path,jones2012layered}, but all of them focus on single-core quantum computers, lacking for a communications perspective. We introduce a generic (i.e no specific qubit technology or interconnect technology is assumed) layered stack that implies multi-core quantum computing by adding the corresponding layers and identifying the communication processes that may be involved. This approach goes beyond adding mere interconnects, encompassing instead communications and computing in a consolidated layered architecture itself --\textit{a là} NoC (Network-on-Chip) \cite{benini2002networks}--. Although there exist some stack proposals extending quantum computers to connected environments, these approaches come from a Quantum Internet perspective, i.e. do not integrate the quantum computation process with communications: they are network stacks rather than computer architecture stacks \cite{pirker2019quantum,wehner2018quantum,dahlberg2019link}.

\begin{figure}
    \centering
    \includegraphics[width=\linewidth]{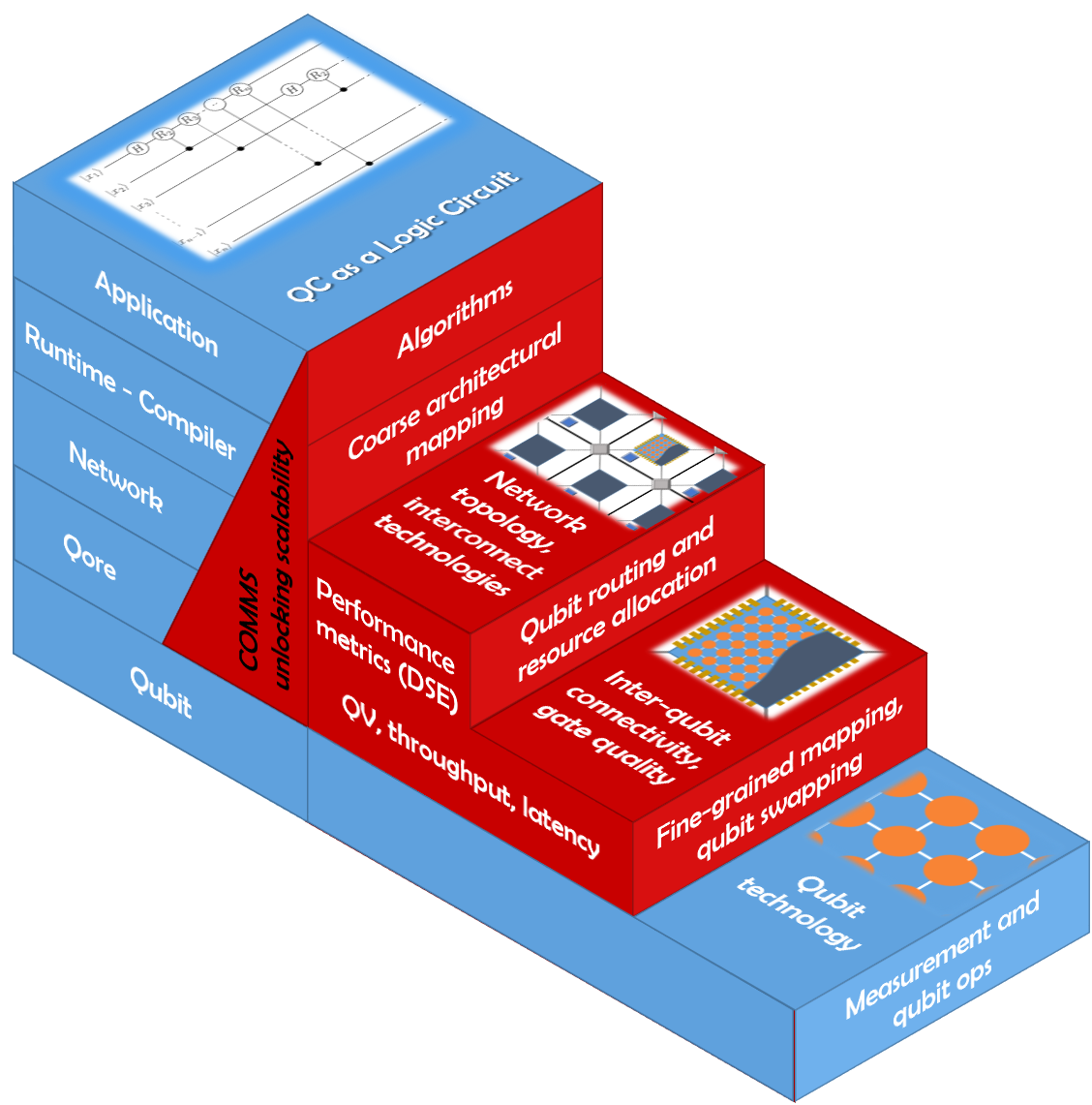}
    \caption{Double joint full-stack layered architecture for multi-core quantum computers}
    \label{fig::NoQCstack}
\end{figure}

The full-stack layered architecture for multi-core quantum computers proposed in this paper can be seen in Fig. \ref{fig::NoQCstack}. In order to represent the different abstractions of the quantum computer at each of the layers, we have included a \textit{stairway} that graphically explains what elements configure that specific layer (on each of the step treads) and its key functions (on the step risers).

The basic layers of the multi-core quantum computer are, then, from logical to physical: Application, Runtime/Compiler, Network layer, Core layer and Qubit layer. A single-core quantum computer would hence have a void Network layer, but would keep the rest of them. Communications (the red ``wedge'' in the figure) are part of the stack in a vertical way, i.e. they affect all the way from the code (which could contain references and optimizations to the qubit communication/distribution, as it is sometimes done in multi-core classical computing), to the qubit movement performed at the Core layer (which is in fact the most basic form of quantum communication). Following is a brief description of each layer:

\begin{itemize}
    \item \textbf{{Application Layer}}. The upper-most layer is composed by the code of the quantum application/algorithm to be run on the quantum computer. In this layer the quantum computer is seen as a Logic Circuit with no reference to limits and architecture for communication among qubits. Nonetheless, the code could include some compiler instructions enabling optimized qubit distribution and instructions execution, as it is already done in multi-core classical computing.
    \item \textbf{Runtime/Compiler Layer}. It is in charge of translating the human-written code to a machine-adapted assembler code (compilation) and coordinate the instructions execution and the coarse architectural mapping (i.e. partitioning of the algorithm among the existing cores, in analogy with the \textit{mapping} process in classical many-core computer architectures), always in pursuit of an optimized processing. Therefore, it needs a closer look to the architecture, knowing about the capabilities and topology of the multi-core quantum computer. However, it is not directly in charge of communicating qubits, something done by the network layer, which is next.
    \item \textbf{Network layer}. This layer (which forms what we could call a Quantum NoC, i.e. a QNoC) may receive some instructions from the compiler regarding qubit movement among cores, but it is fully responsible of selecting the best time and route to do so, as well as to optimize all the inter-core\footnote{\textit{Inter-core} communication involves transferring qubits \textit{among} cores, while \textit{intra-core} communication refers to any type of qubit transmission happening \textit{inside} a core.} communication by reserving resources or preparing qubit movements in advance. It might implement different inter-core topologies (such as all-to-all, star, ring or regular 2D lattices) as well as interconnect technologies (e.g. ion shuttling, qubit teleportation...). Both compiler and network layers \textit{see} the quantum computer as a set of quantum cores (i.e. ``processing units'') connected in a certain topology. Communications are crucial at these layers, as they are ubiquitous in every action performed at this level.
    \item \textbf{Core layer}. In a single-core quantum computer (no network layer), this one represents in fact the whole computer. In any case, the core layer's view is reduced to a set of qubits integrated in a single core capable of interoperate using one and two-qubit gates. It performs the fine-grained qubit mapping inside the core as well as inter-core I/O operations control. Therefore, communications play also here a remarkable role, as qubit movement is the most basic form of quantum communication, and the core needs also to receive input states and send results to other cores. Qubit connectivity inside a core  (encompassing topology and communication type), as well as gate quality are key elements configuring this layer.
    \item \textbf{Qubit layer}. The last one is the qubit layer: the individual qubits, whether they are logical qubits (i.e. a group of qubits acting as a more reliable single qubit) or physical (QEC techniques should be applied instead to handle limited fidelity). No communications are involved, but being the foundation of the whole computer, the performance of this layer is key, as is further explained in the next section. Decoherence processes as well as measurement and gate performance are the main aspects here, and at the same time are highly dependent upon the qubit technology.
\end{itemize} 

This full stack overview of a multi-core quantum computer with built-in communications helps us to show that they play a fundamental role not only in a specific part of it, but in the computer as a whole. Without the communications block (in red), the stack of Fig. \ref{fig::NoQCstack} is unstable. But, the question arises of whether this key block would really unlock quantum computer scalability.


\section{Compressing a quantum computer through models: distilling its essence}
\label{section::vars}

\begin{figure}
    \centering
    \includegraphics[width=\linewidth]{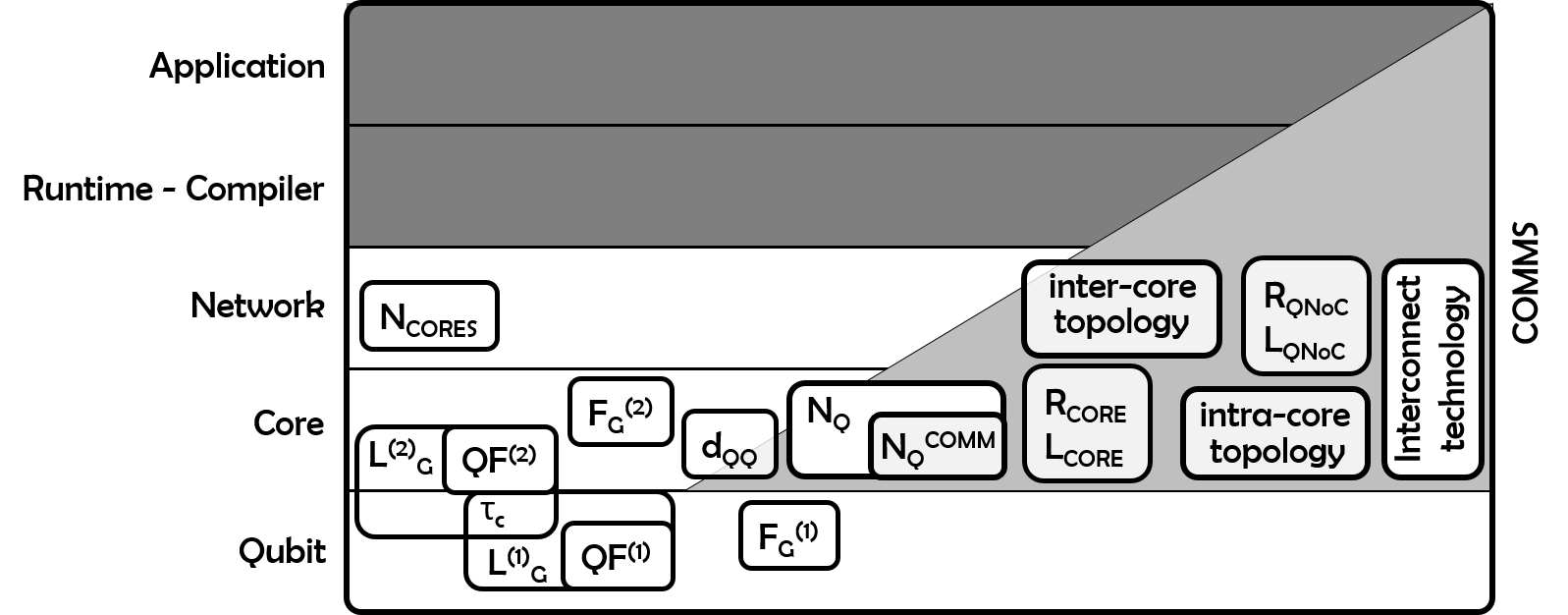}
    \caption{Parameter space, placed within their respective layers in the stack}
    \label{fig::parameters_stack}
\end{figure}

When describing a problem in order to explore the design space with a performance optimization purpose, one must carefully select the elements in the design that are crucial from the point of view of the optimization problem and discard the rest of them, as long as they do not influence the model from the chosen perspective.

In our case, we are exploring whether a multi-core architecture can make the difference in terms of processing power scalability. Therefore, we must particularly take into account the elements of the quantum computer that may be affected by the architecture paradigm (single-core \textit{versus} multi-core).

For this reason, basing upon the layered stack described in the previous section, we will analyze the three lower layers that come into play when considering multi-core architectures, and thus are most affected by intra-core and inter-core quantum communications, namely: qubit, core and network layers (see Fig. \ref{fig::parameters_stack}).

Let us now analyze each layer separately and identify the main parameters to take into account for the subsequent DSE, with special attention to their effect on communications.

\subsection{Qubit layer}

When looking at an individual qubit for the main features that may affect the performance of the quantum computer as a whole, the analysis must take into account the different available technologies as there is no dominating qubit technology yet. Each technology is on a different maturity stage, and presents advantages and disadvantages on the various qubit quality attributes. Proposals for quantum computer implementations include largely developed candidates such as ion traps or superconducting qubits, promising candidates such as quantum dots and other solid-state proposals (such as NV centers in diamond and silicon-based nuclear/electron donor spins), and many others, including photonic and topological qubits \cite{buluta2011natural,resch2019quantum,national2019quantum,van2013blueprint,ladd2010quantum,gerbert2018next}.

Being the most prominent technologies as for now, we do only consider for this work Ion traps, Superconducting qubits, and two types of Solid-state spins: quantum dots and donor spins in silicon.

\textbf{Ion Traps}. In this implementation qubits are represented with the energy levels of electron spins in single ions, that are trapped together in vacuum using electromagnetic fields. It is reliable and presents low error rates (being charged atoms, they can be manipulated with high precision), although operating with them is slow (physically moving ions with electromagnetic pulses) and there are issues scaling beyond around 50 qubits in the same trap.

\textbf{Superconducting qubits}. Using well-known CMOS technology, these qubits are ``artificial atoms'' implemented in superconducting circuits at cryogenic temperatures as LC circuits translated into discrete energy-level systems thanks to the non-linear behavior of Josephson junctions. The information, encoded in different ways (charge, flux or phase qubits) can be electrically controlled via microwaves, voltages, magnetic fields or currents. Leading tech companies such as IBM and Google are betting for this technology, with lower qubit lifetimes and error resistance but faster qubit operation and high flexibility.

\textbf{Solid-state spins}. To avoid the issues with cooling and controlling a large number of atoms in vacuum found in ion traps, several qubit technology proposals are based in ``artificial atoms'' integrated in solid-state host, such as quantum dots (nano-structures of trapped electrons), NV centers in diamond or donor (usually phosphorus atoms) spins in silicon (P:Si). Being still an immature technology, it provides fast operating and long-lived qubits. Moreover, the silicon-based implementations may benefit from the know-how of the global silicon manufacturing experience.

\begin{figure}
    \centering
    \includegraphics[width=\linewidth]{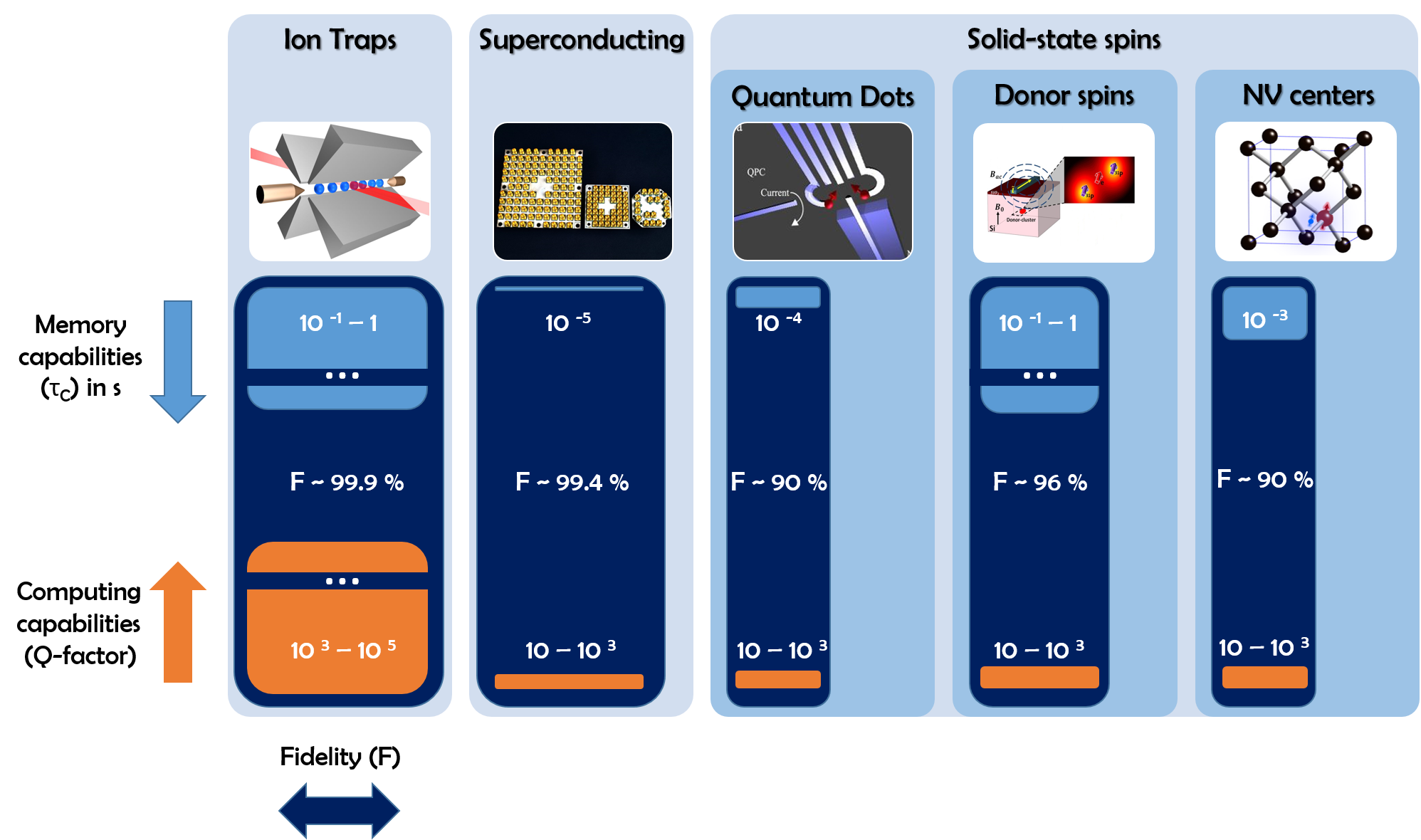}
    \caption{A visual comparison among different quantum technologies. Fidelity percentages correspond to single-qubit gate values. Information source: \cite{resch2019quantum} and \cite{buluta2011natural}}
    \label{fig::qubit_technologies}
\end{figure}

A visual comparison among qubit technologies in terms of their main quality attributes is presented in Fig. \ref{fig::qubit_technologies}.

In the figure, three different characteristics are depicted for each technology: coherence time, quality factor and gate fidelity. Instead of providing figures and metrics, these parameters are represented geometrically: the height (or width) of each box is proportional to the value of the parameter for that technology. As coherence times of both donor spins in Si and ion traps (as well as quality factor in ion traps) are out of scale, their corresponding boxes are represented as a ``discontinued'' surface. In this way, we can easily compare the performance of the different technologies (i.e. of the most recent functional quantum computers developed using each of them).

\subsubsection{\textbf{Coherence time}}
The coherence time $\tau_c$ (i.e. the amount of time the qubit is able to maintain its quantum state unchanged) sets a fundamental limit on the maximum time we can operate and read out the state of the qubit. Thus, $\tau_c$ is proportional to the qubit's memory capabilities. Short coherence times mean short-lived variables (which in turn implies that complex algorithms are not supported), considering that, as stated by the no-cloning theorem, there is no way to replicate quantum information \cite{wootters1982single},\cite{dieks1982communication}: only QEC is able to somewhat extend variables lifetime, generally at a notably high cost in space (qubits used) and time.

Measuring qubit decoherence can be done in two different ways, usually referred to as amplitude damping (or $T_1$) and phase damping (or $T_2$). Amplitude damping is the average transition time from the excited state to the ground state, mainly due to dynamic coupling. Phase damping determines the amount of time the qubit is able to keep a superposition state. Furthermore, if we want to measure the decoherence of a qubit ensemble instead of a single qubit, we use $T^{*}_2$, which takes into account an additional decoherence source, i.e. the uncertainty in the relative phases among different qubits due to the spatial inhomogeneity. This means usually that $T^{*}_2 \leq T_2$, having that $T^{*}_2$ may deviate notably from $T_2$ \cite{wang2005spin,resch2019quantum,almudever2017engineering}. Although these two metrics represent different decoherence phenomena, given that the energy relaxation (amplitude damping) does disturb also the qubit phase, the coherence time $T_2$ is affected by both decoherence processes, and thus it is widely used in the literature as the standard qubit decoherence metric.
Being the main \textit{quality} metric for a qubit, the decoherence times have been continuously improved in the different existing qubit technologies. Although it is not the only challenge for the scalability of quantum computers, the values hitherto reached are still far from allowing qubits to run succesfully representative quantum algorithms without QEC \cite{murali2019full}.

\subsubsection{\textbf{Quality factor}}
The quality factor $QF$ is a parameter derived from the coherence time $\tau_c$ (which, following the literature, we take as $T_2$) and the gate latency $L_G$ (i.e. the time spent in performing a certain quantum operation, such as a Hadamard gate or a CNOT). It is an estimate of the number of gates (quantum operations) that can be applied to a qubit while it contains coherent information (see for instance \cite{resch2019quantum}), and hence is related to the qubit's computing capabilities. The quality factor is computed as:

\begin{equation}
QF= \dfrac{\tau_c}{L_G} = \dfrac{T_2}{L_G}
\end{equation}

It is clear to see the relationship between coherence time and the memory capabilities of that qubit technology: ion traps and quantum dots are technologies well suited to building up quantum memories. The quality factor, however, represents indirectly the computing capabilities of the qubit. A high quality factor allows operating many times on the qubit. Note that it is not necessarily related to large $\tau_c$; a good \textit{memory} qubit may constitute a lower quality \textit{processing} qubit, see e.g. quantum dots. Although this is not part of our main goal for this article, the relationship $\tau_c$ -- $QF$ is quite interesting and may affect quantum processors design and claim for hybrid technologies approaches.

\subsubsection{\textbf{Gate fidelity}}
These two parameters' effect on computer performance is in fact modified by the third parameter, represented in Fig. \ref{fig::qubit_technologies} as the width of each box: the gate fidelity ($F_G$). It is a simplification of the complex quantum error models, and represents how likely a quantum operation will not introduce errors in the system (i.e. inverse to the gate error rate). Low fidelity values will render useless a qubit, no matter how long the coherence time might be, as each operation on the qubit will corrupt its information and affect the entire computation. Although quantum error correction techniques do exist, the uncertainty in low fidelity environments is one of the most limiting parameters of current quantum computers.

\textbf{\textit{Effect on communications}}. The qubit layer is not directly related to any communication process, as we have stated in the previous section. However, being the foundation of the whole computer, it will impose some limits on latencies and qubit rates of upper layers communication processes. This is particularly relevant when we consider that quantum communication is closer to ``transporting physical qubits'' rather than to ``sending quantum information''. The effects of these parameters on the upper layer communication processes are indirect but real, e.g., a short coherence time might cause a qubit state communication to completely fail, if the time-of-travel is longer than $\tau_c$. Long gate latencies (i.e. small quality factors) have a similar effect. Qubits supporting long travels will not withstand too many operations on its already worn quantum state. Finally, low fidelities are equivalent to the inverse of classical communications error rates; the higher the better for accurate quantum information transmission.

\subsection{Core layer}

In the previous subsection we have introduced $\tau_c$, $QF$ and $F_G$. Although the coherence time is directly related to a single qubit, the quality factor and the gate fidelity are usually computed separately for one-qubit gates ($QF^{(1)}, F^{(1)}_G$) and two-qubit gates ($QF^{(2)}, F^{(2)}_G$). Therefore, from the layered stack perspective, \textit{two-qubit fidelity $F^{(2)}_G$} and \textit{two-qubit quality factor $QF^{(2)}$} would in fact become the first parameters of the core layer, as they involve operations among more than one qubit. 

We will use $N_{Q}^{CORE}$ for the \textit{total number of qubits} forming the core. $N_{Q}^{COMM}$ of them will be responsible for interconnecting the core with one or several (identical) modules, if they are integrated in a multi-core architecture and the interconnect technology requires dedicated qubits for communication purposes (else, $N_{Q}^{COMM} = 0$). In the extreme case of a monolithic single-core architecture, no other module exists and hence $N_{Q}^{COMM} = 0$. The interconnection graph might follow a certain topology, whether it is all-to-all, a ring or a regular 2D lattice. Together with the inter-qubit communication technology, it characterizes the \textit{intra-core connectivity}. Finally, the control wiring and qubit technology determine a global minimum \textit{qubit-to-qubit distance} $d_{QQ}$ across the system, which will limit the area occupied by the core and affect the communication latencies.

\textbf{\textit{Effect on communications}}. Inside a quantum core, the most common form of communication is direct swapping, which involves a series of SWAP gates to move the quantum state from any qubit to another one in the same core. The performance of this communication process will be clearly affected by low values of $F^{(2)}_G$ and $QF^{(2)}$, as well as by the topology and the number of qubits $N_{Q}^{CORE}$; a large processor with an uneven topology may need on average longer travels. In other types of communication, such as qubit shuttling, the inter-qubit spacing will determine the travel distance (and duration). In any case, it is of interest for the analysis to derive a mean \textit{intra-core communication latency $\bar{L}_{CORE}$ and qubit rate $\bar{R}_{CORE}$}.

\subsection{Network layer}

At this layer, we can see the whole quantum communications (QNoC) perspective. Parameters interesting to our analysis include \textit{inter-core connectivity} (both in terms of topology and interconnect technology) and the \textit{number of cores} in the processor $N_{CORES}$.

\textbf{\textit{Effect on communications}}. Depending on these parameters we will obtain different values for mean \textit{inter-core communication latencies and qubit rates} ($\bar{L}_{QNoC}$ and $\bar{R}_{QNoC}$, respectively). Other design decisions such as the qubit routing algorithm and resource allocation implementations complete the set of variables that will affect communication processes in our environment.

        

\begin{table}[!t]
\renewcommand{\arraystretch}{1.9}
\caption{Notation and symbol definitions}
\label{table::notation_table}
\centering
\begin{tabular}{|p{0.2\linewidth}|p{0.6\linewidth}|}
        \hline
        $\tau_c$ & Coherence time\\
        \hline
        $L_G^{(1)}, L_G^{(2)}$ & One and two-qubit gate latency\\
        \hline
        $F_G^{(1)}, F_G^{(2)}$ & One and two-qubit gate fidelity\\
        \hline
        $QF^{(1)}, QF^{(2)}$ & One and two-qubit quality factor\\
        \hline
        $N_Q$ & Total number of qubits in a quantum computer\\
        \hline
        $N_{CORES}$ & Number of cores in a multi-core quantum computer\\
        \hline
        $N_Q^{CORE}$ & Total number of qubits in a quantum core\\
        \hline
        $N_Q^{COMM}$ & Number of qubits dedicated to inter-core communications in a core\\
        \hline
        $N_Q^{MAX}$ & Upper limit of total number of qubits that can be integrated in a quantum computer\\
        \hline
        $N_Q^{LIM}$ & Upper limit of total number of qubits that can be integrated in a single quantum core\\
        \hline
        $d_{QQ}$ & Global minimum qubit-to-qubit distance\\
        \hline
        $L_{CORE}$ & Quantum communication latency in a core\\
        \hline
        $R_{CORE}$ & Qubit rate in a core\\
        \hline
        $L_{QNoC}$ & Quantum communication latency in inter-core communications (QNoC)\\
        \hline
        $R_{QNoC}$ & Qubit rate in inter-core communications (QNoC)\\
        \hline
    \end{tabular}
\end{table}

\section{On a Design Space Exploration for double stack communications-enabled quantum computers}
\label{section::DSE}

\begin{figure*}
    \centering
    \includegraphics[width=0.8\linewidth]{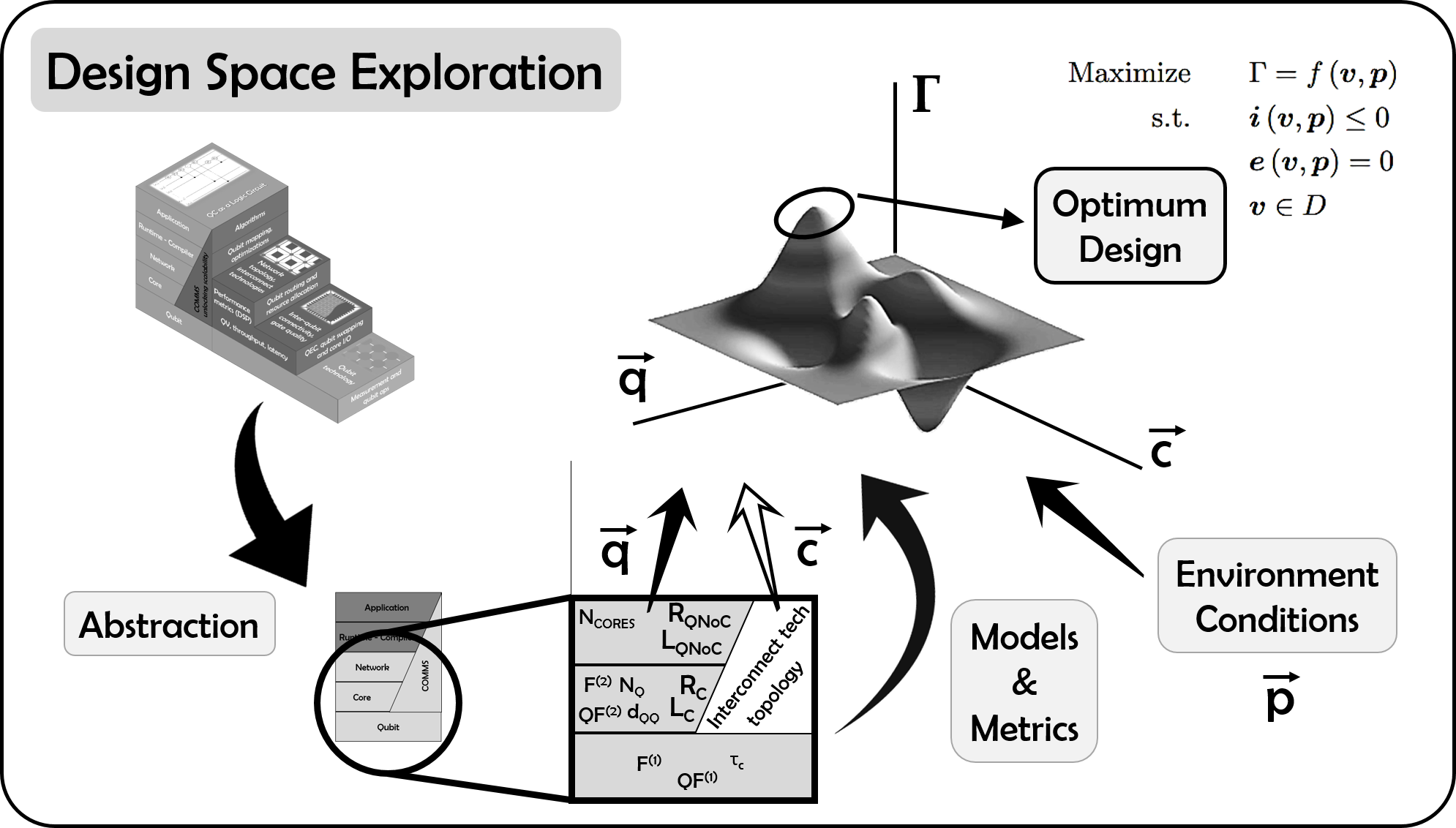}
    \caption{A Design Space Exploration for Multi-Core Quantum Computers}
    \label{fig::DSE_in_a_fig}
\end{figure*}

Now that we have described the parameters that model the proposed approach of a double stack communications-enabled quantum computer, it is time to present the methodology we will use to solve the problem we are facing (DSE), and depict its application to our context (see Fig. \ref{fig::DSE_in_a_fig}).

\subsection{The Design Space Exploration technique}

DSE is a structured design methodology that allows to optimize a system maximizing a given cost function --or Figure of Merit (FoM)-- based on some parameters of interest \cite{kang2010approach,gries2004methods}. Like any other structured design process, this optimization relies on modeling the interdependencies among the different performance metrics and the variables describing the system. This modeling process might include analytic/theoretical expressions, behavioral models, computer-based simulations, or their zone-wise combinations.

It is important to note that DSE is used to \textit{design}, not just to \textit{optimize} (performance metrics optimization is in fact just one of the DSE use cases -- DSE is also useful for rapid prototyping or system integration with no need for analytical metrics \cite{gries2004methods}). Whatever the design problem is, if the analysis is correctly prepared, the DSE analysis will not blindly look for ``the extreme-case-highest-performing scenario''. Rather, the main virtue of DSE is to be able to consider system-wide trade-offs and different metrics that may also affect the design optimality. For example, a DSE analysis of a network deployment will not optimize the average throughput of the entire network, but will take into account deployment costs and qualitative characteristics such as network reliability or flexibility. DSE achieves this by letting the designer to concurrently sweep all the open variables in the design space --instead of ``manually'' tweaking them in a one-by-one approach--, and to consolidate several performance/cost/qualitative metrics into a single merit figure $\Gamma$, which is then optimized.

DSE can be applied to any design problem, from a very specific single-variable local problem to a System Architecting Problem (SAP), such as the one we are facing. In either case, DSE optimization can be generically formulated as in Eq. \eqref{eq::generic_opt_problem}, where $\Gamma$ is the merit figure, $\boldsymbol{J}\left(\boldsymbol{v},\boldsymbol{p}\right)$ is the vector containing the different optimization metrics considered, $\boldsymbol{v}$ is the vector containing the decision variables that determine the system design, and $\boldsymbol{p}$ the vector of fixed parameters that specify the environment/scenario the system is placed on. The equalities in vector $\boldsymbol{e}$ and inequalities in vector $\boldsymbol{i}$ must be satisfied, in order for the solution to remain within the feasible domain $D$.

\begin{subequations}
    \label{eq::generic_opt_problem}
    \begin{align*}
        \max_{\boldsymbol{v}} & \qquad \Gamma = f\left(\boldsymbol{J}\left(\boldsymbol{v}, \boldsymbol{p}\right)\right) \tag{\ref{eq::generic_opt_problem}}\\
        \text{s.t.} & \qquad \boldsymbol{i}\left(\boldsymbol{v}, \boldsymbol{p}\right) \leq 0\\
        \text{ } & \qquad \boldsymbol{e}\left(\boldsymbol{v}, \boldsymbol{p}\right) = 0\\
        \text{ } & \qquad \boldsymbol{v} \in D
    \end{align*}
\end{subequations}

The advantages of this methodology are thus threefold:
\begin{itemize}
    \item Exploring the entire design space without being limited by the ``intuition'' and designer's previous experience that might hinder the way to the optimal (but maybe not intuitive) solution.
    \item Providing not just a single optimal analytical solution, but rather design trends and guidelines extracted from the exploration.
    \item Being valid also for early design decisions, when there are no analytical models or computer simulations for the performance metrics of the system.
\end{itemize}

\subsection{Applying DSE}

Applying DSE to our specific optimization problem implies identifying each one of the generic components of the formulation above in the multi-core quantum computing environment. That is, choosing one or more performance metrics (describing the computational power of the resulting design) to evaluate the whole multidimensional design space, which is defined using a specific set of variables and parameters. 


We have already carried out a first description of the system parameters and open variables in our system design (layer by layer, see Section \ref{section::vars}). That is, they integrate the vectors $\boldsymbol{p}$ and $\boldsymbol{v}$; depending on the aspects to be analyzed, a given element will be considered in the analysis as a fixed parameter (part of the scenario under study, i.e. part of $\boldsymbol{p}$) or as an open variable (being part of $\boldsymbol{v}$ the DSE will explore different values for that element looking for the one that may contribute to the optimal solution). As an example, let us take $N_{CORES}$. If $N_{CORES} \in  \boldsymbol{p}$, we have fixed its value for our exploration, otherwise, $N_{CORES} \in  \boldsymbol{v}$, and DSE will sweep over different values of $N_{CORES}$. Accounting for our computing-communications double-stack approach, $\boldsymbol{v}$ is divided in two different axis: $\boldsymbol{q}$ (variables coming from the pure computing stack, such as gate latencies or fidelities) and $\boldsymbol{c}$ (those that are part of the communications).

In the Eq. \eqref{eq::generic_opt_problem}, the merit figure $\Gamma$ appears as a function of the vector $\boldsymbol{J}$, which at the same time depends on the system variables $\boldsymbol{p}$ and $\boldsymbol{v}$. Clearly, the definition of the components of $\boldsymbol{J}$ (i.e. the different performance/cost metrics) and the aggregation of them into $\Gamma$ (via the $f$ function) are crucial for a proper DSE
. A generic form for the function $f$, in the form of a normalized weighted sum, is shown in Eq. \eqref{eq::merit_figure}, in which the metrics to be maximized are grouped in the numerator and the ones to be minimized in the denominator.

\begin{equation}
    \label{eq::merit_figure}
    \Gamma = \dfrac{\prod_i{w_i\cdot J_i}}{\prod_j{w_j\cdot J_j}}
\end{equation}

Choosing a complete set of metrics $\boldsymbol{J}$ is not a trivial task, and even more in the case of quantum computing. Therefore, we will spend some time reviewing the actual status of performance metrics and benchmarks in quantum computing and reasoning on the best fitting merit figure $\Gamma$ for our problem.

\subsection{Selecting the performance metrics that perform best}
\label{subsection::metrics}

When facing an optimization problem, it is crucial to accurately define the function to be maximized or minimized. Using numerical/analytical tools to search for the optimum design solution for a given problem implies proper modelling, that is, translating the problem and the optimization goal into a figure, a metric. This modelling has already been done in the previous sections for the proposed system architecture. However, modelling the goal is just as important, as an incomplete optimization metric will render biased the entire design or at least lead us to sub-optimal solutions.

DSE performs an optimization on the merit figure $\Gamma$. For the sake of the DSE merit figure's completeness, it should aggregate metrics on performance, cost and qualitative attributes \cite{araguz2018optimized}. Let us analyze these metrics in the context of quantum computing (the interested reader may consult \cite{resch2019benchmarking} for a complete review on this topic).

Quantum computing is, at this moment, a pure \textit{research} field. 
Because of that, measuring the progress of quantum computing is, right now, more about calibrating its \textit{maturity} and overcoming individual obstacles rather than benchmarking computational power \cite{blume2019metrics}.



Few system-level metrics have been defined for quantum computers. Rather, qubit- and gate-level metrics related to their error performance (such as coherence time, gate fidelities and latencies) have been widely used, and several methods for measuring them have been developed (e.g. direct fidelity estimation and state gate set tomography). 
The reason for this is simple. Integrating several qubits and gates into a single \textit{quantum computer} is quite recent, and the research has been mostly focused only on one and two qubit demonstrations.

Algorithms used as benchmarks in experimental quantum computing are, thus, adapted to its current limited size and capabilities, and interdependencies among single-qubit errors makes extrapolating benchmark results from small computers to larger ones impractical. Because of that, 
existing metrics and benchmarks are still far from capturing all the limitations of a quantum computer design\cite{blume2019metrics,resch2019benchmarking}. Nevertheless, their validity for driving progress on present research is beyond reasonable doubt. In fact, quantum tomography \cite{d2003quantum}, miniature versions of key algorithms \cite{scully2001quantum, fowler2004implementation} or randomized benchmarking \cite{emerson2005scalable,knill2008randomized} are extensively accepted for technology demonstrations, and new approaches are continuously proposed and improved \cite{bishop2017quantum,mccaskey2019quantum,erhard2019characterizing,mills2020application}.

However, consensus on a generic qubit-technology-agnostic metric is essential in order to establish a common framework for a fair comparison among existing prototypes and technologies. Although specific to NISQ systems, a metric that has recently generated interest in the research community is the proposal stemming from IBM's quantum research team \cite{bishop2017quantum,cross2019validating}. It aims at defining a single figure, the so-called \textit{Quantum Volume}, that aggregates the most important factors affecting performance, such as number of physical qubits, number of gates that can be operated before errors make results useless, qubit connectivity and number of operations that can be run in parallel. The QV is computed using the largest random square circuit the quantum computer is able to execute successfully. Authors in \cite{blume2019volumetric} highlighted the limitations of QV and generalized the concept by presenting the \textit{Volumetric Benchmarks}. These metrics are the first to give an architecture-neutral measurement, aiming at evaluating the useful amount of quantum computing performed by a given system.

Applying these proposals as a metric for the DSE analysis of multi-core quantum computers is not straightforward. The elements considered in these metrics do not include communication latencies or qubit rates at different layers. They do not differentiate either the communication time and the computational time. 

Having this in mind, we could conclude that an optimal performance metric for multi-core quantum computers should be:
\begin{itemize}
\item \textbf{Communications-oriented}. Adding communication overheads and other considerations related to the multi-core nature of the proposed double-stack architecture claims for accurate modelling of communication processes inside the quantum chip and possibly designing specific benchmarks.
\item \textbf{Adaptive}. The non-universality and expiration time of current benchmarks (including QV) implies $\Gamma$ should evolve along with quantum computation to avoid misleading designs. Its definition should take into account qualitative attributes.
\item \textbf{A multidisciplinary effort}. A full-fledged analysis requires very refined models (for elements as diverse as qubit cross-talk, fidelity degradation, quantum communication technologies...) and a complete definition of $\Gamma$, something that cannot be obtained without the collaboration of all the fields involved (physicists, material engineers, electrical and computer engineers...).
\end{itemize}

\subsection{A Figure of Merit for Multi-Core Quantum Computers}

Although we aim at developing a complete FoM with exhaustive models, for this introductory paper, we have used intuitive yet useful performance metrics and models, which are aggregated into the FoM $\Gamma$ shown in Eq. \eqref{eq::first_gamma_def}, i.e. it is a preliminary example for illustrative purposes on the usage of DES for multi-core quantum computer design. For that reason, not all of the previously described architectural parameters have been included and the application of DSE is very straightforward, without leveraging all of its advantages. As a behavioral model, this first attempt suffices for showing all the possibilities that DSE has to offer.

\begin{subequations}
    \label{eq::first_gamma_def}
    \begin{align}
        &\Gamma = \dfrac{w_{Qb} J_{Qb} \cdot w_{QF} J_{QF} }{w_{F} J_{F} \cdot w_{I} J_{I} \cdot w_{C} J_{C}} \tag{\ref{eq::first_gamma_def}}
    \end{align}
\end{subequations}
where:
\begin{subequations}
    \begin{align*}
    &w_{Qb}, w_{QF}, w_{F}, w_{I}, w_{C} \in (0,1]
    \end{align*}
\end{subequations}

This definition accounts for different errors and overheads that may synthesize the effect of:

\textbf{i)} the exponential increase of \textit{computational power} when incrementing the number of qubits in the system:
\begin{subequations}
    \begin{align*}
        &J_{Qb} = 2^{\tilde{N}_{Q}} - 1
    \end{align*}
\end{subequations}
where:
\begin{description}
\item $N_{Q}$ is the total number of qubits in system.
\item ${\tilde{N}_{Q}}$ is a normalized version in order to have $J_{Qb} \in [0,1]$.
\end{description}

\textbf{ii)} the \textit{qubit quality}, using the quality factor:
\begin{subequations}
    \begin{align*}
        &J_{QF} = QF = \dfrac{\tau_c}{L_G}
    \end{align*}
\end{subequations}

\textbf{iii)} the degradation of the \textit{aggregated fidelity} as we concentrate more qubits in the computer:
\begin{subequations}
    \begin{align*}
        &J_{F} = 2 - F^{N_{Q}}
    \end{align*}
\end{subequations}
where $F$ is the 2-qubit fidelity.

\textbf{iv)} the \textit{cross-talk and other physical impairments} derived from integrating many qubits in the same chip/core, defined as the mean qubit density per core (multiplied by the qubit-to-qubit disturbance $\epsilon_I$). When the qubit density exceeds the threshold $N_{q}^{MAX}$, the aggravation is degradated by a penalty factor:
\begin{subequations}
    \begin{align*}
        &J_{I} = 1 + \dfrac{\epsilon_I N_{Q}}{N_{CORES}} \cdot  {\left(\dfrac{\mathcal{H}\left(N_{Q} - N_{Q}^{MAX}\right) \cdot N_{Q}}{N_{Q}^{MAX}}\right)}^3
    \end{align*}
\end{subequations}
where:
\begin{description}
\item $N_{Q}^{MAX}$ is the aggregated maximum of qubits that may be integrated into $N_{CORES}$ (the number of cores in the processor).
\begin{subequations}
    \begin{align*}
        &N_{Q}^{MAX} = N_{Q}^{LIM} \cdot N_{CORES}
    \end{align*}
\end{subequations}
\item $N_{Q}^{LIM}$ is the maximum number of qubits that may be integrated into a single core without incurring in severe crosstalk.
\item $\mathcal{H}(x)$ is the Heaviside step function:
\begin{subequations}
    \begin{align*}
        &\mathcal{H}(x) =  \begin{cases}
                                   0 & \text{if $x<0$} \\
                                   1 & \text{if $x \geq 0$}
                            \end{cases}
    \end{align*}
\end{subequations}
\end{description}

\textbf{v)} the \textit{communications overhead} when using more than one core, defined as an exponentiation of the overhead factor $\epsilon_C$ (error rate increase due to communications overhead when adding a core to the system). For the sake of simplicity, this generic metric accounts for the fidelity degradation in core-to-core communications without specifying the communication technology employed:
\begin{subequations}
    \begin{align*}
        &J_{C} = 2 - \left(1-\epsilon_C\right)^{N_{USED}}
    \end{align*}
\end{subequations}
where:
\begin{description}
\item $N_{USED}$ is the number of cores that contain active qubits (i.e. qubits that are being used in the given configuration):
\begin{subequations}
    \begin{align*}
&N_{USED} = \left(N_{Q}/N_{Q}^{LIM}\right)
    \end{align*}
\end{subequations}
\end{description}




In a brief intuitive summary, the model could be described in the following terms. First, the higher number of qubits we are capable of use in our computer, the better the performance ($J_{Qb}$), with an exponential dependence. A higher quantum technology quality factor $QF$ will also contribute to the performance of the system ($J_{QF}$). Same happens with fidelity ($J_F$), the higher, the better. Of course, using more qubits lowers the aggregate fidelity (which we model as an exponential function). Another aspect to consider is the effect of integrating a large number of qubits in a single core ($J_I$). We model that using a simple linear relationship when below the limit of qubits that may fit in a given system ($N_{Q}^{MAX}$). When we integrate a higher number of qubits, the error increases exponentially. While we are using a number of qubits below that maximum, the performance improves exponentially, and then it saturates. Finally, the communications overhead is to be accounted ($J_C$). When using more than one core, the communication processes (be it quantum teleportation, ion shuttling...) that are needed to operate among qubits placed in different cores might be costly and set a limit for the multi-core approach. They depend upon the mapping algorithm and the communication technology employed. In this first model, this overhead is represented using an exponential dependence on $N_{CORES}$.

\section{Results}
\label{section::results}

In order to visualize and clarify the possibilities of DSE, we now present the results of a first DSE analysis on multi-core quantum computers looking for answers to some of the most interesting questions in quantum computing: How will the quantum computer scale in number of qubits? Will multi-core approach unlock the current monolithic single-core quantum computers' scalability bottlenecks? Does the inter-core communications technology affect the performance of multi-core quantum computers? How the existing qubit technologies compare as candidates for multi-core quantum computing? 






\subsection{Scalability analysis}

\begin{figure*} 
    \centering
  \subfloat[\label{1a}]{%
       \ifthenelse{\boolean{comms_saw}}{\includegraphics[width=0.32\linewidth]{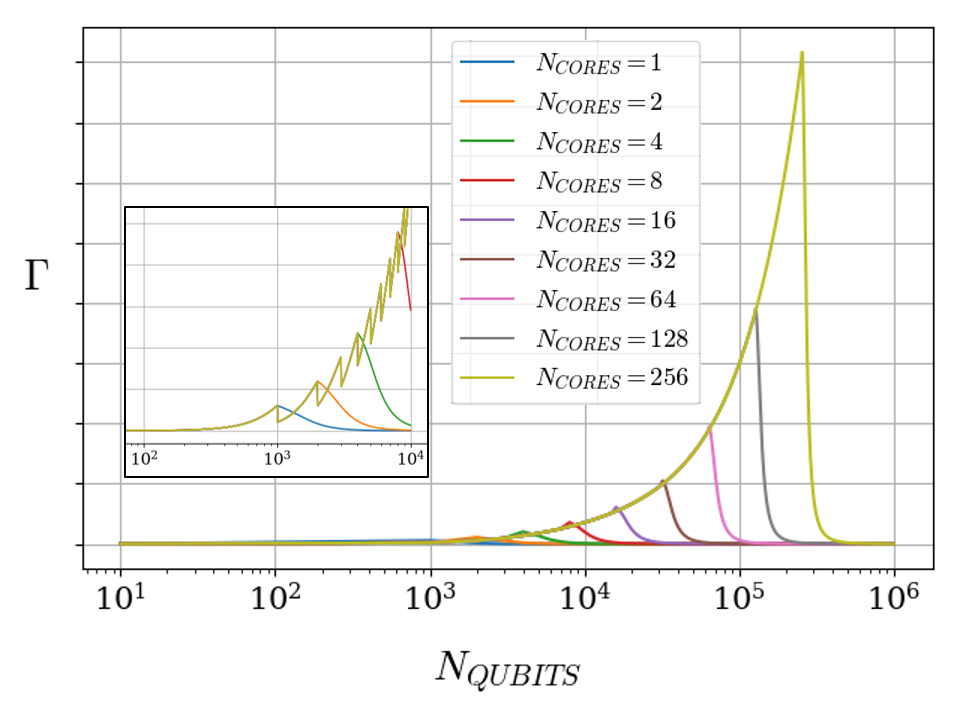}}{\includegraphics[width=0.32\linewidth]{name/fig/scal_plot_with_zoomin.png}} }
    \hfill
  \subfloat[\label{1b}]{%
        \ifthenelse{\boolean{comms_saw}}{\includegraphics[width=0.32\linewidth]{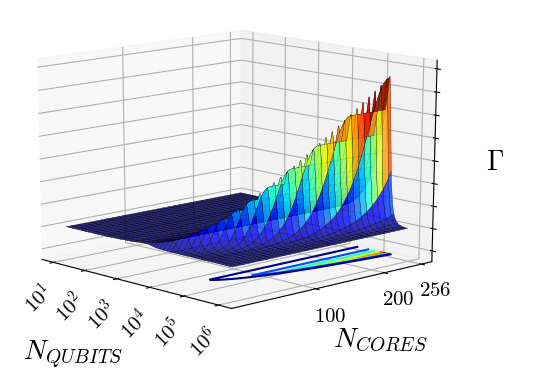}}{\includegraphics[width=0.32\linewidth]{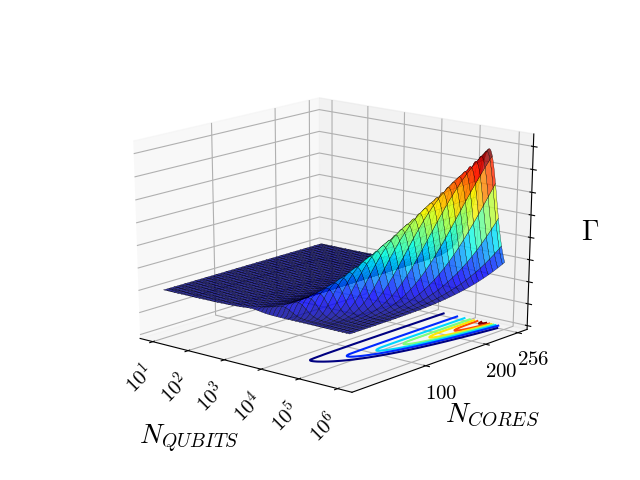}} }
    \hfill
  \subfloat[\label{1c}]{%
        \ifthenelse{\boolean{comms_saw}}{\includegraphics[width=0.32\linewidth]{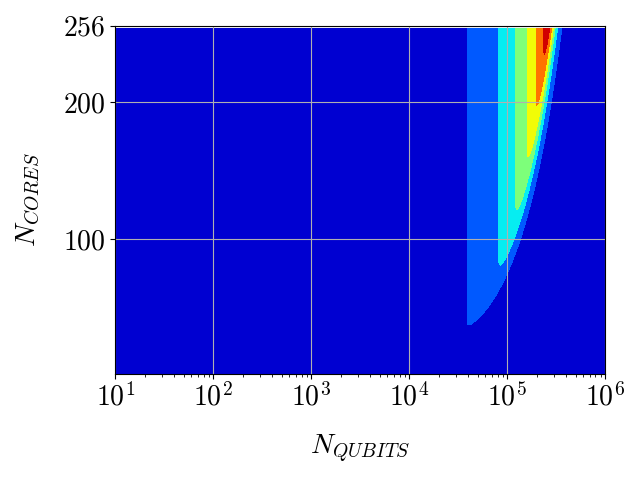}}{\includegraphics[width=0.32\linewidth]{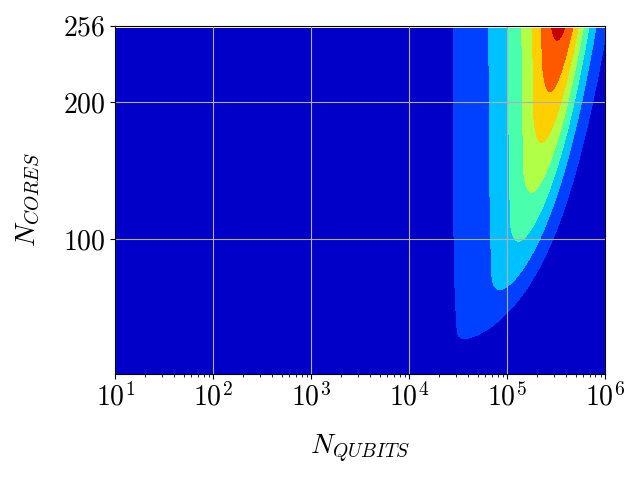}} }
  \caption{\textbf{Scalability analysis} (a) Quantum computer's overall performance is plotted against the number of qubits used in the system, for several configurations in terms of number of cores used. Qubit operations' fidelity $ F= 99.9 \%$, $\epsilon_C = 5 \%$, $\epsilon_I = 0.1 \%$, and $w_i = 1, \forall i$. A maximum number of 1000 qubits are allowed to be integrated in the same core. Observe the communications overhead effect on performance. (b) and (c) Performance analysis when varying both the number of qubits and the number of cores in the quantum computer. The isolines in the plot let us know different configurations that provide the same performance.}
  \label{fig::synthetic_scal_plots} 
\end{figure*}

The main concern of the present article, and thus the main result expected from the DSE analysis, is to determine whether the multi-core approach will effectively supersede current NISQ computers and enable their scalability into large quantum computers. And of course, if we take into account a specific scenario or a given set of requirements (i.e. having fixed $\boldsymbol{p}$, e.g. core-to-core communication latency, gate fidelity upper-bound, existing (or predicted) coherence time $\tau_c$ range, etc.), we will be able to study the evolution of the design performance ($\Gamma$) when sweeping over the number of cores $N_{CORES}$ from 1 (single-core \textit{traditional} quantum processor) to tens or hundreds of them. This is called \textit{scalability analysis}. This type of analysis allows to detect scalability trends and bottlenecks, hence obtaining design guidelines on quantum computer configurations for an optimal performance.

Using the assumptions and models presented in the previous section, we present such analysis in Fig. \ref{fig::synthetic_scal_plots}, where we show the $\Gamma$ values for a wide range of quantum computer configurations. We have selected $N_{Q}$ and $N_{CORES}$ as our vector $\boldsymbol{v}$, fixing the rest of elements. That is, the exploration is restricting the scalability to the increase in number of qubits and cores, and the relationship among them. The fixed parameters ($\boldsymbol{p}$) have been set to realistic values for ion trap technology (the most evolved yet) taken from \cite{resch2019quantum} and \cite{national2019quantum}, that is, single-qubit gate fidelity $F = 99.9 \%$, gate latency $ L_G = 5.4 \cdot 10^{-7}$, coherence time $\tau_c = 2 \cdot 10^{-1} s$, $\epsilon_I = 0.0001$, $\epsilon_C = 0.05$, maximum number of qubits per core $N_{Q}^{LIM} = 1000$, and FoM weights $w_i = 1, \forall i$ (no specific metric is set as a prevailing target).

In the leftmost plot, a single-core quantum computer is compared to several multi-core configurations, for a total number of qubits $N_{Q}$ in the system varying from $10$ to $10^6$ qubits. For each configuration, the performance ($\Gamma$) follows a peaky bell-shape trend, with a maximum close to $N_{Q} = N_{Q}^{MAX}$ (the optimal configuration for that number of cores). Trying to integrate more than $N_{Q}^{LIM}$ qubits in a single core causes a steep degradation of performance. The single-core processor is clearly exceeded by multi-core approaches. This first model does not capture a realistic communications overhead when using many cores, thus the performance is monotonically increasing in $N_{CORES}$. Of course, this might change when considering refined communication models and mapping strategies, that may work better with a low number of cores and suffer from worse communications overheads. In the zoom-in, the saw-like profile in the performance curve can be clearly seen. Whenever the optimal qubit distribution requires another core to be used (if available in the configuration), the extra comms overhead causes a steep fall in performance. This implies that the configuration with more cores is not always the best performing one. The center and right-most plots contain a complete input variables sweeping, with $N_{Q}$ varying from $10$ to $10^6$ qubits, and $N_{CORES}$ from $1$ to $256$. The isolines in the plot let us know different configurations that provide the same performance, e.g. $10^5$ qubits in a 200-core quantum computer performs the same as twice as many qubits using less than half of the cores. Observe that the more cores are present in the system, the narrower is the performance curve, that is, a low number of cores guarantees good relative performance for both low and high number of qubits. That may collapse for a sufficient number of qubits, where no matter the number of cores, the performance may be the same or worse (due to communications overheads).
 
Using this simple model, we can clearly draw three main conclusions: \textit{i)} the QNoC approach is promising as a scalability enabler, \textit{ii)} for every multi-core quantum computer configuration there exists an optimal working range ($\Gamma$ over a certain minimum threshold), and \textit{iii)} the $N_{Q}^{LIM}$ parameter clearly constrains the performance of the configuration and thus we should consider it as a fundamental design variable. With more accurate data and models, we postulate that this type of analysis will effectively accelerate and optimize the research on Quantum Computing.
 
 \subsection{A Quantum Technology Gap Analysis}
\begin{figure*} 
    \centering
    \subfloat[\label{2a}]{%
       \includegraphics[width=0.5\linewidth]{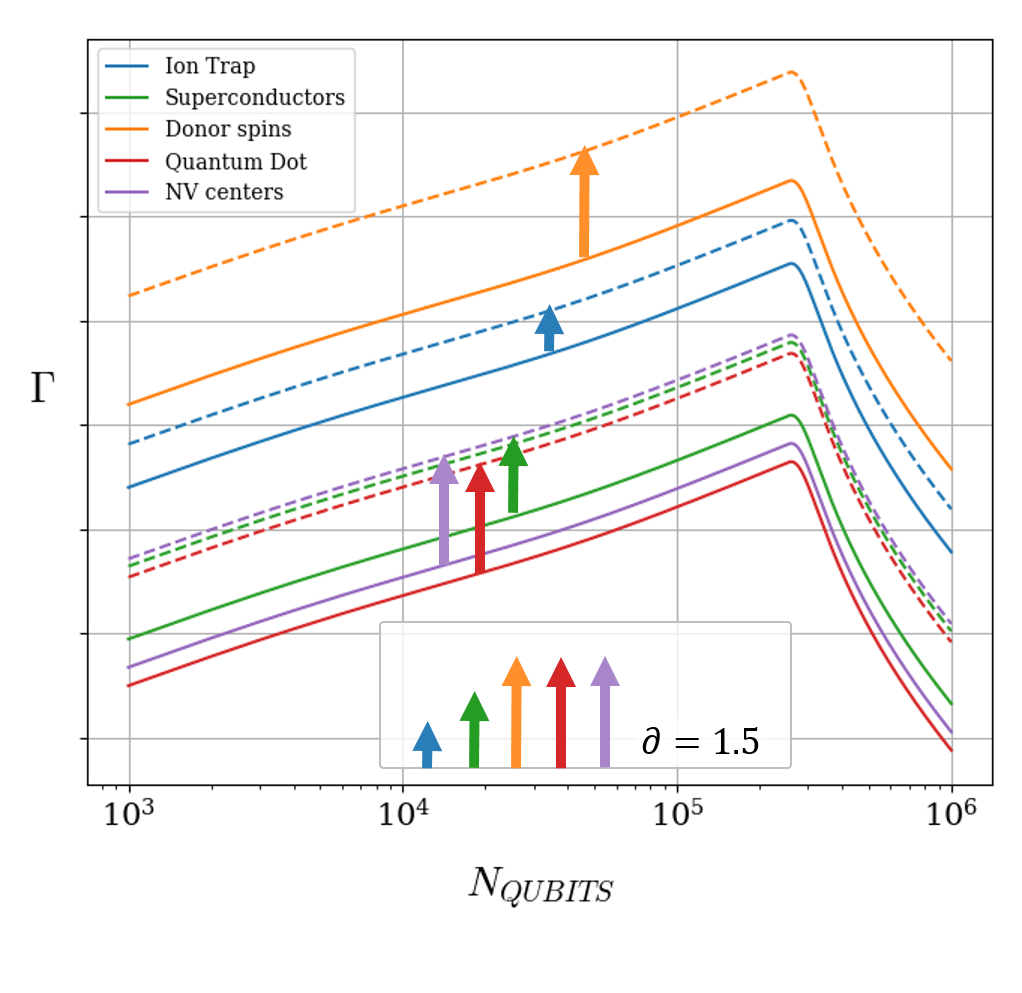} }
    \subfloat[\label{2b}]{%
        \includegraphics[width=0.49\linewidth]{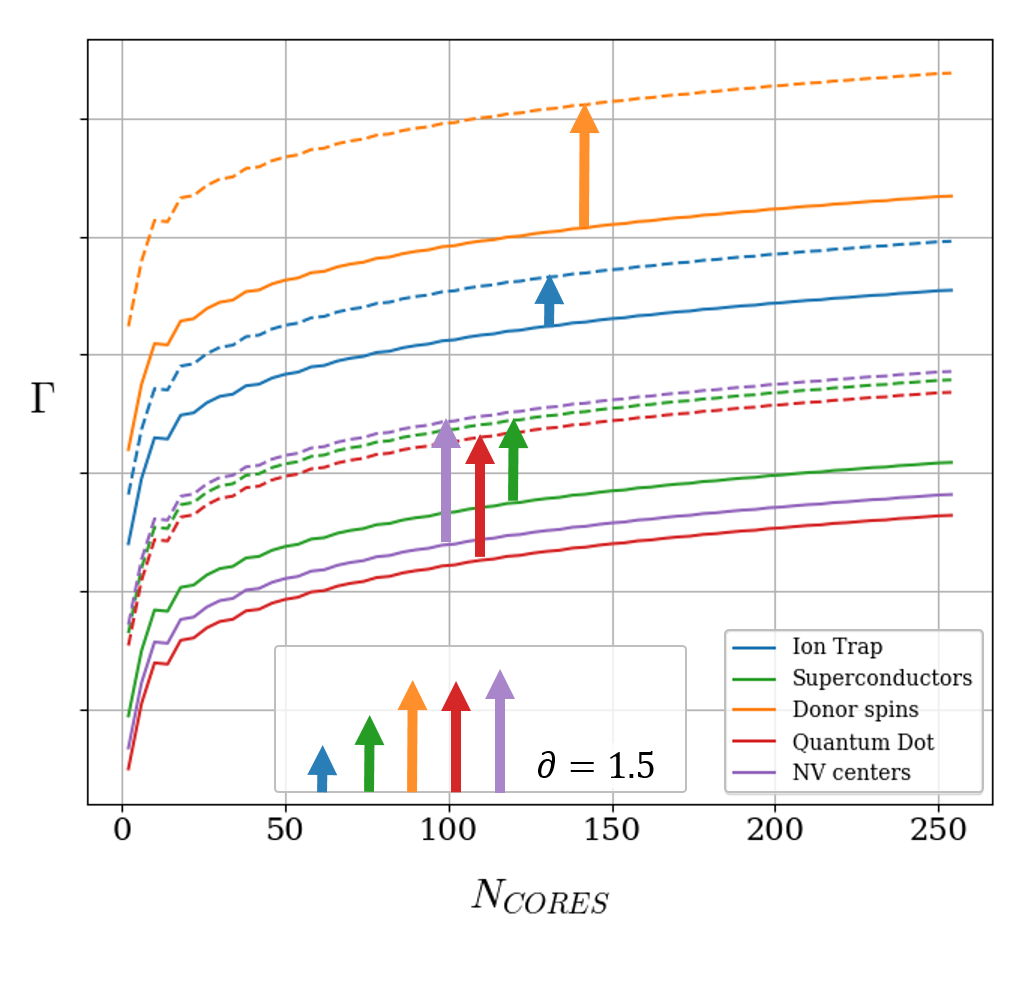} }
    \caption{\textbf{Quantitative qubit technology gap analysis} (a) Quantum computer's overall performance $\Gamma$ for existing qubit technologies is log-plotted against the number of qubits used in the system. For each technology, current parameters are used to draw the solid line, and a $\delta = 1.5$ evolution/improvement on them is reflected in the dashed line. The number of cores is set to 256, and the rest of parameters are fixed as in previous figure. The validity of the conclusions drawn from this technologies comparison depends exclusively on the accuracy of the models used (b) In this case, the performance is plot against the number of cores in the system. The performance value assigned to a given number of cores corresponds to the peak performance of that quantum computer configuration (i.e. the number of qubits is set to be the optimal for every value of $N_{CORES}$).}
    \label{fig::synthetic_delta_plots_qubits_vs_cores} 
\end{figure*}

 However, DSE might be applied for analysis that go further than stating the benefits of the application of QNoC to quantum computers. It may also help Quantum Engineers to optimize the effort and investments in quantum computing. Therefore, with accurate models, we are able to explore the whole design space in order to focus research and experiments only on the most promising materials, technologies or parameter ranges. As an example, we have performed a simple Quantitative Technology Gap Analysis, i.e. a performance comparison of the existing qubit technologies and their evolution in the next years that opens a window to the future, letting us to know which technologies may provide higher profitability after a certain research investment. To do so, a common performance metric, such as the previously defined golden metric $\Gamma$, is needed, in order to establish a common ground for a fair comparison.
 
 In Fig. \ref{fig::synthetic_delta_plots_qubits_vs_cores} the comparison among all the qubit technologies (and their projected performance) is shown, both for varying $N_{Q}$ and $N_{CORES}$. Each technology is represented using the three most representative physical parameters: $\tau_c$, $L_G$ (aggregated in $QF$) and $f$. Using actual measurements retrieved from \cite{resch2019quantum} and \cite{buluta2011natural} as the parameters baseline, $\delta$ represents a correlative delta improvement on the three of them. In the case of quality factor $QF$, $\delta$ represents a constant proportional improvement (i.e. $QF_{\delta} = QF * (1 + \delta)$ ). The fidelity is set to improve asymptotically to 100 \%. Under the used assumptions and models, we could already draw some conclusions on the comparison of existing technologies, e.g. donor spins in Si seem to be the best performing technology, with still much room for growth.
 
 \begin{figure*}
    \centering
    \includegraphics[width=\linewidth]{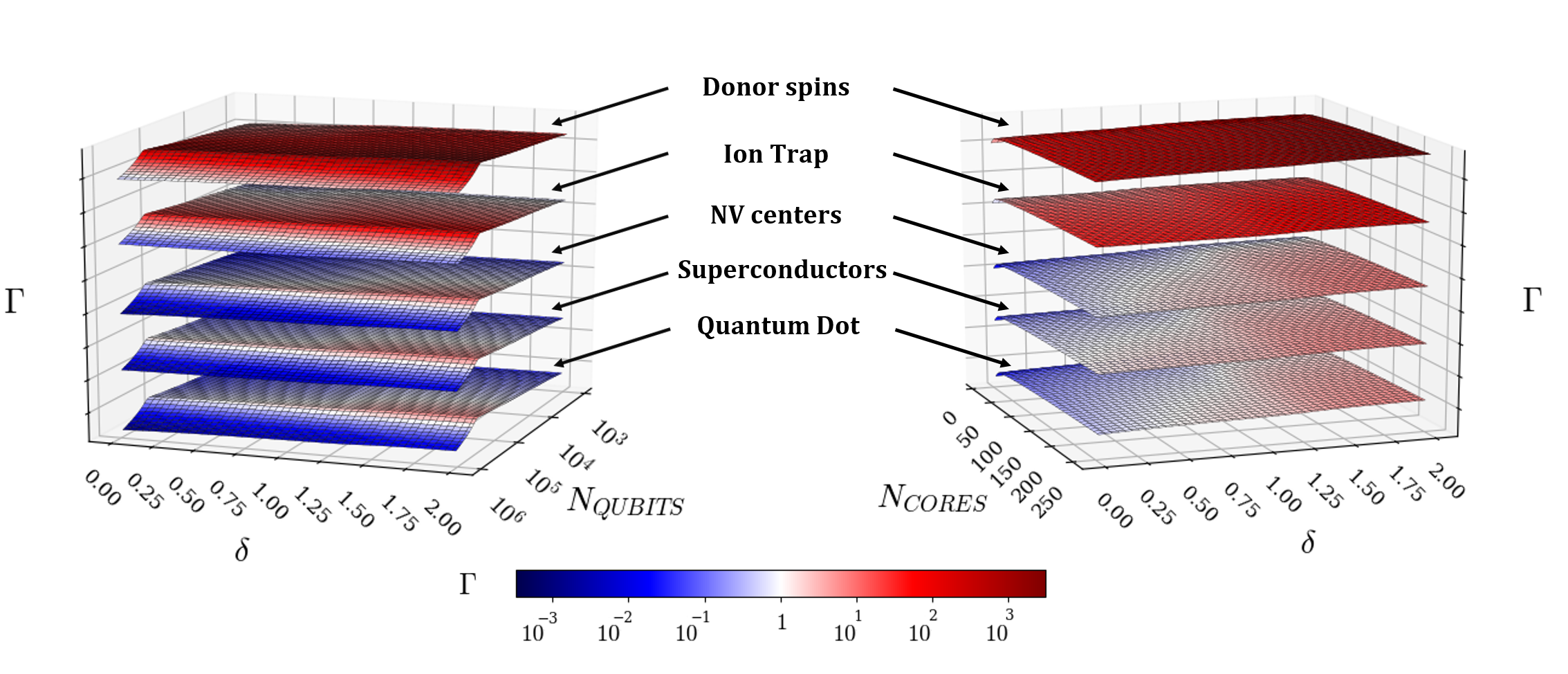}
    \caption{QTGA log plot extended for a wide range of $\delta$, for existing qubit technologies (technologies are offset in Z axis for the sake of clarity, while keeping the color mapping for $\Gamma$ values), sweeping the number of qubits and the number of cores in the system. Observe that Donor spins performance evolution with $\delta$ is really promising, when compared to the other technologies, which in comparison remain almost plain. When exploring for increasing number of qubits, the number of cores is set to 256, while for varying number of cores the number of qubits is determined as the one providing peak performance. The rest of parameters remain the same as in Fig. \ref{fig::synthetic_scal_plots}.}
    \label{fig::synthetic_3D_delta_plots_all_techs}
\end{figure*}

 \begin{figure*}
    \centering
    \includegraphics[width=\linewidth]{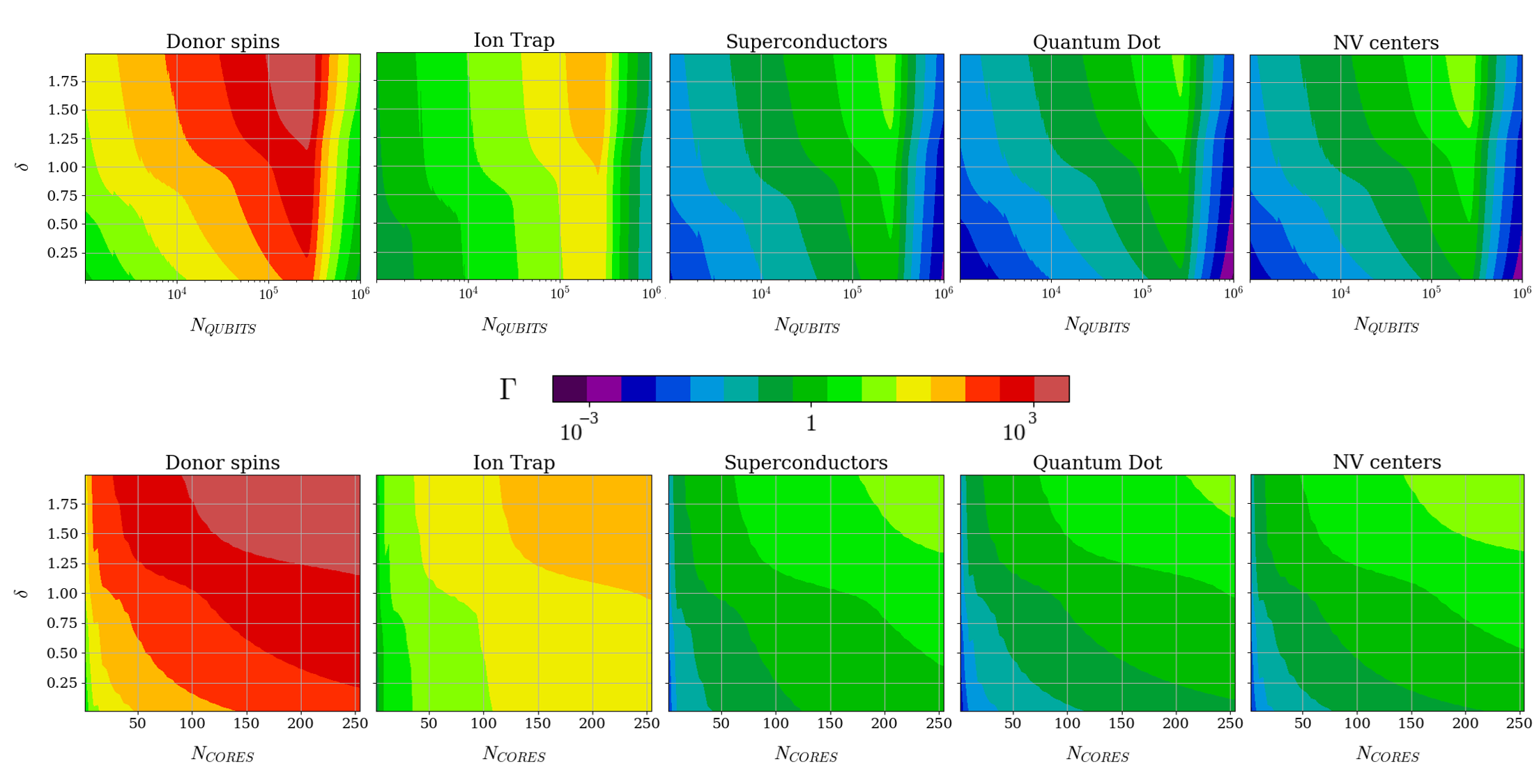}
    \caption{Performance plot extended for a wide range of $delta$, for existing qubit technologies, sweeping the number of qubits in the system. Observe that performance does not increase linearly with $delta$, nor does it behave the same for different technologies. Equivalent performance can be obtained with a notably lower number of qubits if the technology is improved.}
    \label{fig::synthetic_delta_plots_all_techs}
\end{figure*}
 
 It is important to highlight the non-linear behavior of performance improvement with $\delta$. See Figs. \ref{fig::synthetic_delta_plots_all_techs} and \ref{fig::synthetic_3D_delta_plots_all_techs}. In Ion Trap, for instance, in the interval $\delta = (0.5,0.6)$ the performance escalates exponentially. It corresponds to fidelity values going from 99.999 \% to 99.9999 \%, that is, investments to make improvements in this interval will pay off abundantly. Moreover, we can conclude that, even with a $100\%$ improvement, only Ion Trap can place performance on a range similar to that of solid-state based quantum computer. Observe also that a $\delta$-evolution on qubit technologies does not only improve peak performance, but also the optimal operational margin. For instance, $30.000$ Ion Trap qubits ($\delta = 1$) integrated in a 256-core quantum processor provide the same performance as $200.000$ (5 times more) current Ion Trap qubits. A similar behavior is present in all technologies. When studying the effect of varying the number of cores in the system, we observe as before a clear performance improvement in $\delta$. However, it is to be remarked the ``conversion factor'' that we find of a $\delta$-evolution and number of cores needed to achieve the same performance. Observe that, for all technologies, a $\delta = 0.6$ suffices for having a 70-core processor matching a processor more than twice as large in number of qubits and cores.

\section{Discussion}

The results presented in the previous section allow us to foresee not only the promising effects of applying multi-core architecture to quantum computers, but also the possibilities of DSE analysis for quantum computer design. With more accurate data, FoM and models, we postulate that drawing this sort of conclusions as outcomes of DSE will effectively accelerate and optimize the research on quantum computing.

\subsection{One analysis to rule them all} 


We have just seen two different applications of DSE: exploring scalability and comparing qubit technologies. However, we are able to explore the design space in many ways. We could, for instance, determine which is the best performing design in terms of interconnects technology, intra- and inter-core topology, or even qubit technology. This last analysis (determining the best performing qubit technology) would provide an answer to one of the most currently exciting questions in quantum computing. To avoid unfair comparisons, the analysis could be complemented with a ``maturity-related'' metric included in $\Gamma$. The ultimate FoM might be based on existing performance metrics, such as QV, while also take into account the considerations we have made at the end of Section \ref{subsection::metrics}. 

In this way, using DSE with accurate models, we expect to be able to state results such as ``among the existing qubit technologies, basing upon model predictions, the technology \textit{X} is the most promising for building multi-core quantum computers in high-latency environments'', or ``a lattice topology provides the lowest latency for quantum teleportation-based intra-core networks'', and specially ``the multi-core quantum computer approach performs better than the monolithic single-core designs when more than \textit{N} qubits are needed, and/or intra-core latencies are lower than $L_{max}$, and/or coherence times $\tau_c$ are in the range $(t_L, t_H)$''.

\subsection{On communication needs}
The potential of DSE may help us in overcoming several engineering challenges that arise in current quantum computer design. However, we shall not forget the main developement bottleneck, i.e. current quantum computer architectures are not scalable. Therefore, though there are many valuable elements that should be considered when working on next-generation quantum computing, the key barrier to be removed is the scalability issue and the DSE analysis must be directed into that. The approach proposed in this paper, involving a QNoC interconnecting multiple quantum cores, essentially suggests that introducing communications will be the turning point for quantum computers into the decisive production phase.

In Sections \ref{section::statement} and \ref{section::vars} we have seen that communications are present both in the intra- and inter-core qubit-to-qubit operations. This implies communications performance greatly affects system-wide efficiency. From swapping qubit information inside a core, to distributing an entire algorithm through several cores and controlling the QNoC flow, communications must be carefully considered in the DSE analysis.

Apart from generic parameters such as latency or qubit rate, topology/connectivity and the technology chosen for every link present in the architecture play a key role in the design. For instance, the efficiency of qubit swapping for transmitting quantum information in a single core in superconducting qubits might be very different to that provided by qubit shuttling for ion trap qubits, and each of them will have a working optimal range in terms of number of qubits or fidelity requirements. Another example, choosing teleportation for inter-core communication implies introducing an overlaid classical network connecting the cores and providing the architecture with bell-pair preparation nodes. And the last one, using a fully-connected inter-core network may lower the average communications latencies and error rates, but might imply a prohibitive control overhead and cost in terms of communications-dedicated qubits per core.

Therefore, together with specific compilers for multi-core architectures and bell-pair provisioning protocols, specific and high quality models for quantum communications are needed in order to be able to carefully tailor this essential element to the quantum computer design.

\section{Conclusions and future work}
\label{section::conclusions}

In this paper we have presented a double joint full-stack layered architecture for quantum computers that introduces communications in a multi-core approach (Quantum Network on Chip) as a scalability enabler for performance-unlocked quantum computing. We have also introduced a system-wide optimization proposal (using Design Space Exploration) that might facilitate once-for-all design guidelines unifying the still separated design technologies into a consolidated solution with optimal technologies and parameters for every situation. This will definitively allow to happen all the unprecedented advances expected in application fields such as pharmacology, internet security and big-data analysis that we expect, as well as those that we cannot even imagine.

An initial DSE analysis has been presented using intuitive basic models that help us to imagine the world of possibilities that DSE enables, via a scalability analysis and the very first quantitative qubit technology gap analysis. With the present and future work of all the quantum community in the models we need to improve and perfect the DSE (e.g. fidelity or coherence time dependencies on gate and qubit technologies, models relating qubit communication error rates with inter-network topologies or number of qubits per core, etc...) we will be able to elaborate future analysis including quantum computers benchmarks comparison, an improved qubit technology gap analysis and provide design guidelines, with a special emphasis on self-specification of QNoC.






\bibliographystyle{IEEEtran}
\bibliography{IEEEabrv,bib/main}
\end{document}